\documentclass[]{spie}  

\usepackage[]{graphicx}

\newcommand{\beq}{\begin{equation}}
\newcommand{\eeq}{\end{equation}}
\newcommand{\bqa}{\begin{eqnarray}}
\newcommand{\eqa}{\end{eqnarray}}

\newcommand{\erf}[1]{Eq.~(\ref{#1})}

\newcommand{\dg}{^\dagger}

\newcommand{\smallfrac}[2]{\mbox{$\frac{#1}{#2}$}}
\newcommand{\half}{\smallfrac{1}{2}}
\newcommand{\bra}[1]{\langle{#1}|}
\newcommand{\ket}[1]{|{#1}\rangle}
\newcommand{\ip}[2]{\langle{#1}|{#2}\rangle}
\newcommand{\sch}{Schr\"odinger}

\newcommand{\hei}{Heisenberg}

\newcommand{\cu}[1]{\left\{{#1} \right\}}

\newcommand{\an}[1]{\left\langle{#1}\right\rangle}

\title{Optical coherence and teleportation: \\ Why a laser is a clock, not a quantum
channel.} 

\author{Howard M. Wiseman
\skiplinehalf School of Science,
Griffith University,  Brisbane,
Queensland 4111 Australia}

\authorinfo{Further author information:
E-mail: H.Wiseman@griffith.edu.au; 
 Full affiliation: Centre for Quantum
Computer Technology, Centre for Quantum Dynamics, School of Science,
Griffith University}

\begin{document} 
  \maketitle

\begin{abstract}
It has been argued [T. Rudolph and B.C. Sanders, Phys. Rev. Lett. {\bf 87}, 
077903 (2001)] that continuous-variable quantum
teleportation at optical frequencies has not been achieved because the
source used (a laser) was not `truly coherent'. Van Enk, and Fuchs [Phys.
Rev. Lett, {\bf 88}, 027902 (2002)], while arguing against Rudolph and Sanders,
also accept that an `absolute phase' is achievable, even if it has not
been achieved yet. I will argue to the contrary that `true coherence' or
`absolute phase' is always illusory, as the concept of absolute time on a
scale beyond direct human experience is meaningless. All we can ever do is
to use an agreed time standard. In this context, a laser beam is
fundamentally as good a `clock' as any other. I explain in detail why this
claim is true, and defend my argument against various objections. In the
process  I discuss super-selection rules, quantum channels, and the
ultimate limits to the performance of a laser as a clock. For this last
topic I use  some earlier work by myself [Phys. Rev. A {\bf 60}, 4083
(1999)] and Berry and myself [Phys. Rev. A {\bf 65}, 043803 (2002)] to
show that a Heisenberg-limited laser with a mean photon number $\mu$ can
synchronize $M$ independent clocks each with a mean-square error of
$\sqrt{M}/4\mu$ radians$^2$.  
\end{abstract}


\keywords{Quantum teleportation, optical coherence, quantum optics, 
laser linewidth, quantum channel, super-selection rules, clock
synchronization.}

%
\section{INTRODUCTION}
\label{sect:intro}  

Recently, there has been an interesting controversy
\cite{RudSan01,Wis01b,EnkFuc02a,EnkFuc02b,RudSan01b,NemBra02,Wis02b,HeyBjo03} 
in the quantum information and quantum optics community about the status
of the first continuous variable quantum teleportation (CVQT) experiment \cite{Fur98}{}.
It began when Rudolph and Sanders (RS) published a letter entitled 
`Requirement of Optical Coherence for Continuous-Variable Quantum
Teleportation' \cite{RudSan01}{}. In it they argued that CVQT  was not
achieved in Ref.~\cite{Fur98}{}, and in fact that it cannot be achieved
using a laser as a source of coherent radiation. They base their argument
on their claim that a laser is {\em not} a source of coherent radiation,
in the sense that the output of a laser is not a coherent state, 
but an equal mixture of coherent states with all possible phases.

It is true that
any good quantum optics text (e.g. 
Refs.~\cite{SarScuLam74,WalMil94}) will
show that the stationary 
solution of the master equation for a laser
yields a state which is 
an equal  mixture of coherent states of all
possible phases:
\beq
\rho_{\rm ss} = \int \frac{d\phi}{2\pi} \ket{|\alpha|
e^{i\phi|}}\bra{|\alpha| e^{i\phi|}},
\eeq
where $|\alpha|^2 = \mu$ is the
mean photon number of the laser. 
 As RS correctly point out, following M\o lmer \cite{Mol96}{}, this 
solution can also be interpreted as a mixture of
number states, as
 \beq
 \rho_{\rm ss} = \sum_n
e^{-\mu}\frac{\mu^n}{n!}\ket{n}\bra{n}.
\eeq
Thus, they argue, the
description in Ref.~\cite{Fur98} is invalid 
because it relies upon the
`partition ensemble fallacy' \cite{KokBra00}{}. That is, its analysis 
is carried using one partition 
of the ensemble  (into coherent states)
because it would not be valid in another partition (into number 
states).

An important motivation for RS is as follows \cite{Rudpc}{}. If one 
were to use the number state partition to describe the teleportation
experiment, one would find that the entanglement  necessary to do the
teleportation is contained not in the conventionally recognized
``entanglement resource" of twin-beam squeezing.  Rather it would be 
a property only of the entangled resource {\em plus}  an extra laser 
beam continuously shared between the two parties. Conventionally the laser is
taken to be in a coherent state. If the phase of the 
coherent state is assumed to be unknown then this is equivalent to  
using the coherent-state
partition. In this formulation the extra 
laser beam is regarded simply as
establishing a time-reference.  In 
the number-state partition, it appears
as part of the entangled 
resource, which suggests that it is really
another quantum channel, 
which is continuously open. Since quantum
teleportation is supposed 
to take place in the absence of a quantum
channel, this brings the 
validity of the experiment into doubt. 

In this paper I will argue that the position of RS is not tenable by 
any practicing physicist or engineer, and that there are good reasons 
to say
that sharing a laser beam is simply establishing a time-reference. That
is, {\em a laser is a clock, not a quantum channel}.  Van Enk and Fuchs
(EF)\cite{EnkFuc02a,EnkFuc02b}  have been the most active opponents to RS,
but I will argue that their position is as flawed as that of RS, and for
essentially the same reason. 

This paper is structured as follows. I will
present my basic argument, that a laser is a clock, in Sec.~2. This is
drawn largely from my presentation in Ref.~\cite{Wis02b}{}. The remaining
sections (some of which are also based on Ref.~\cite{Wis02b}) are devoted
to counter-arguments, and their flaws, as I see them. Some of these
objections are published by others, some are unpublished (that is,
personal communication), and some are hypothetical. I will use my
responses to these objections to freely explore a number of 
interesting issues related to this question. In particular, I will 
discuss super-selection rules and their relation to conservation 
laws, quantum versus classical channels, how good a laser is as a 
clock. The latter is
quantified by the result that a quantum-limited laser with a mean photon
number $\mu$ can synchronize $M$ independent clocks with a mean-square
error of $\sqrt{M}/4\mu$ radians$^2$.  

\section{My Argument}

There is nothing wrong with the mathematics of RS. However, their letter suffers
from a deep conceptual problem:  they assume the existence of an absolute
time or phase standard at optical frequencies. 
The reason they do so is that they do not wish to claim that CVQT at 
optical frequencies is impossible in principle. If they had claimed 
this then at least they would
have established a consistent position, but also  they would have 
also have displayed the practical absurdity of it. That is because if 
one insists on 
rejecting coherent optical sources, then one must discard much
else besides.

To allow for the possibility of optical CVQT, RS  must allow that 
the production of coherent states
of light is possible in principle, by some means other than a laser.
Indeed
they say that `We therefore assert that genuine CVQT 
requires {\em
coherent devices}, that is, devices capable of 
generating true coherence,
and these are not a feature of {\bf 
current} [my emphasis] CVQT
experiments.' If this were true then RS 
would have some basis for
criticizing at least the formalism of 
Ref.~\cite{Fur98}{}. 
It is the point of my argument here to show that 
even in principle there are no
devices that can generate `true 
coherence' any better than a laser.

Again emphasizing that their argument relies upon the possible 
existence
of `true coherence', RS say that they `are of course {\em 
not}  asserting
that production of coherent states of light is 
impossible: basic quantum
electrodynamics
shows that a classical oscillating current can produce
coherent
states.' The first problem with this claim is that, as far as we know, the universe is 
quantal, not classical. There is no reason for
believing in {\em classical} electrical currents.  Nevertheless, 
a suitable quantum current would generate a 
coherent state of light to an
arbitrarily good approximation, so I 
will not belabour this point. 

A more 
serious problem is that it is not possible to produce even such a
quantum current, in a sense that would pass the `truth' test of RS.  
The natural oscillators
at optical frequencies are the electrons in
atoms.
Without coherent light, we could still make the atom `ring'  by
`striking' it (with a free electron, for example). 
However, for this
oscillation to  have a definite 
phase the time of the collision would have
to be known to an
accuracy less than an optical cycle, of order
$10^{-15}$s.
Otherwise one would have to average
over all possible phases
and one would be left with exactly
the same problem as with the laser.

Achieving this accuracy would require 
a clock which ticks faster than
$10^{15}$s$^{-1}$. Presumably this is 
the sort of `technical challenge'
that RS mention in the context of 
producing coherence by making a
measurement on the gain medium of a 
laser. But to call it a 'technical
challenge' is to miss the point 
that it simply moves the problem back
another step. 
How could we be sure that the clock has a definite phase?
Perhaps it is fixed relative to another clock, but then we can ask the
question again of 
that second clock. And so {\em ad infinitum}.

It is thus clear that the `true coherence' as meant by 
RS is impossible to
produce technologically. 
It cannot mean coherent relative to any clock,
because 
their argument attacking the phase of a laser can be equally used
to attack 
that of any clock. Therefore, if it means anything, `true
coherence' 
must mean coherent relative to an absolute time standard for the 
universe. Since 
such a hypothetical absolute time standard can never
be measured, 
I would maintain that it is meaningless, and with it the idea of 
`true coherence' in the sense of RS. 

With no absolute time, all we
can ever do is to use an agreed time 
standard, and measure phase relative
to that. In this context, a laser beam is as good a 
`clock' as any other.
The electrodynamic coupling Hamiltonian 
$\propto {\bf p}\cdot{\bf A}({\bf x})$ allows, in principle, the 
laser clock 
to influence the motion of a
charged particle with position and 
momentum ${\bf p}$ and ${\bf x}$. In
physical terms, when the 
electric field points up, an electron is
deflected down, and vice 
versa. This can be used to synchronize the laser
with any 
material clock.

Once material particles are synchronized with
the laser field, they 
could then  be synchronized with  any other clock
based on any gauge 
boson field, via  analogous coupling Hamiltonians
\cite{Ryd85}{}. 
This includes gravitational and nuclear force fields, but
not the Higgs field or the ``second quantized" field of any composite
bosons such as mesons or Hydrogen atoms. Material oscillations of course
must involve a force field, and hence also involve the oscillation of 
a gauge boson field of some sort. The fungibility of the time 
standards based on different gauge boson fields makes all of these 
time-keepers
equivalent, and justifies calling them 
`clocks'.  This is not an empty
definition, as the other (non-gauge) 
boson fields, are not fungible time
keepers. For example, an 
atom laser beam cannot be a clock in this sense.
Its phase can only be
defined or measured relative to another atom laser
beam of the same 
species \cite{Wis97}{}. 

Fundamentally, the difference
between (for example) an atom laser and an optical 
laser is that the atom laser 
field arises from the quantization of identical particles, not the 
quantization of a field. This explains the difference in 
the way the field appears in a
Hamiltonian. In coupling Hamiltonians 
a gauge field can appear
linearly. By contrast, the field 
describing particles always appears
bilinearly, and in conjugate 
pairs; if one particle is destroyed, another
is always created. 
If one is familiar with Feynman   diagrams, then the
difference can be 
understood from the fact that every vertex involving a
particle has 
{\em two} lines for particles, but {\em one} 
(wiggly) line
for the gauge field \cite{Ryd85} . In 
the standard model, the particles
(in this sense of the word) are 
fermions, but the consequences is seen in
the non-fungibility of the phase of the bosonic field for composite particles such 
as in an atom laser.

The above arguments
lead inevitably to the conclusion that 
in quantum optical experiments
there is no necessity to consider, even hypothetically, any 
time-keeper
beyond 
the laser which serves as a phase reference.  No other clock is
superior in any fundamental sense. Now by definition a laser beam is
perfectly 
coherent relative to itself (ignoring experimentally negligible
phase 
and amplitude fluctuations, and transverse mode incoherence). 
Thus, the phase reference laser beam in the 
teleportation experiment  {\em is}
in  a coherent state. It is not an unknown coherent state, it is the
coherent state $\ket{\sqrt{\mu}}$, with zero phase (relative to 
itself).  There is no process that will make a coherent state 
 in any stronger
sense, and no need for for any stronger sense. There 
 is no unmet need for
optical coherence in continuous-variable 
 quantum teleportation.

\section{Objection: We experience absolute time}
 
The first obvious
objection that might be voiced is that  
as conscious beings we experience
the flow of absolute time
directly, and so give it meaning. Admitting  
the validity of this 
temporal experience (which does not go without saying
\cite{Peg91}), this argument nevertheless cannot work to establish
optical coherence.  We cannot
simply look at a clock ticking every
$10^{-15}$s and verify that it 
has a definite 
phase, because we cannot
perceive anything in $10^{-15}$s. 

My rebuttal applies not only to optical
frequencies. Experiments show that our perception of time
has a resolution
in the range of tens or even hundreds of 
milliseconds \cite{Pen90}{}.
Thus {\em we cannot 
establish the absolute phase of any oscillator
of frequency greater than a few tens of Hertz.} At higher frequencies 
we can
only establish the phase of one oscillator relative to another 
oscillator. This conclusion is not altered by oscillations obtained by 
frequency combs or $2^{n}$-upling \cite{Jon00}{}. The timing of the 
zeros of the
highest harmonic 
can be no more accurately defined than that of 
the fundamental.

From personal observation, discussions between physicists
about the existence of `true coherence' or `mean fields' 
often end with
one party waving an
arm up and down, intimating that by so moving an
electric charge,  a mean field would be produced. But that appeal fails
precisely when the frequency is too fast for us consciously to move 
any part of our body at that frequency. That is, it again fails for 
frequencies greater than a few tens of Hertz.
 
 There are many qualitative
differences between the way radiation is generated, or detected, 
across
the spectrum. But the only location for a fundamental 
(if fuzzy) dividing
line in frequency between 
absolute and relative phase, or between `true
coherence' 
\cite{RudSan01} and `convenient fictions' \cite{Mol96} 
would be between oscillations we can 
observe directly  and those we cannot. If
this division is unpalatable 
that is because it relies upon the notion of
absolute time. By abandoning this ill-conceived  notion,  the  
dividing line between supposed `truth' and supposed `fiction' 
disappears.

\section{Objection: A laser is not ``really'' in a coherent state}

One
would be entitled to be puzzled by the fact that I began by 
seeming to
accept that a laser at steady state is described by a mixture, as in
Eqs.~(1) or (2), but ended by claiming that it is in a pure state
$\ket{\sqrt{\mu}}$. 
This certainly troubled Fuchs \cite{Fuchs:pc}{}. I
will give my 
resolution below, but before that I should say a few words
about the 
published opinions of Fuchs, with  van Enk,
\cite{EnkFuc02a,EnkFuc02b} on RS~\cite{RudSan01}{}. 
My understanding of
their papers is that they are an explicit 
calculation 
(which can be done
analytically because they ignore the laser gain 
process) 
showing how a
laser beam, without an absolute phase, can 
function as a phase reference.
Specifically, they show 
how the phase information can be distributed and
how there is no harm in regarding the phase as real. This is essentially
the same point originally made by M\o lmer, that the laser phase is a
`convenient fiction' \cite{Mol96}{}. The analysis of EF 
gives a rigorous information-theoretic definition of `convenience', 
in terms of the
quantum de Finetti theorem \cite{EnkFuc02b}{}. However, 
it could well be
questioned whether convenience, however rigorously 
defined, is 
sufficient
for a dispensation from the ban on the `partition 
ensemble fallacy'.

The real lesson that should have 
been drawn from the analysis of RS is 
that quantum teleportation can be, and should have been, defined 
operationally,
so that the reality of the laser phase would have been 
irrelevant. The
reality of the laser phase can nevertheless be 
defended, as I have done,
on the grounds that it is no less real than any other 
phase.  However,
this is not the position of EF. 
Instead, they base their entire analysis
on an unquestioning acceptance of Eqs.~(1) and (2) \cite{irony}{}. 
I call this acceptance unquestioning because they 
follow RS in assuming that
there is such a concept as `true coherence', and that this is something
that a laser lacks by virtue of being in a mixture of the form of Eqs.~(1)
and (2).  

 Indeed, in Ref.~\cite{EnkFuc02a}{}, EF 
 directly imply that
there is a time standard more real than that 
offered by the laser itself:
\begin{quote}
However, recent developments
\cite{Jon00} may make it
possible to compare the phase of
an optical light beam directly to the
phase of a microwave
field. Using this technique, the only further
measurement
required \ldots is a measurement of
the {\bf absolute phase}
[my emphasis] of the microwave field, which 
is possible
electronically.
This measurement would create an
optical coherent state from a standard
laser source for the
first time.
\end{quote} 
The implication is that an
electronic measurement somehow makes the phase real, which it was not when
it was an optical phase.  As I have argued above, there is no reason to
regard the laser phase as any less real than any other phase. There is
nothing gained in, and no need for, appealing to any other clock in order
to say that 
the laser is in a coherent state. 
 
Thus I can finally state
the resolution of the puzzle with which I 
began this section. 
In general,
any state of an oscillator is meaningless without a 
specification of the
time reference (i.e. clock) by which it is 
defined. (The exception is an
energy eigenstate.) A state may have an 
ill-defined phase relative to one
clock, but be perfectly coherent 
relative to another one. In particular,
if the laser itself is the 
clock then by definition it is coherent with
respect to itself so 
that it should be described by a pure state
$\ket{\sqrt{\mu}}$ of 
zero phase.  It is ironic that one (Fuchs) whose
mantra is that 
quantum states are states of knowledge (see for example
Ref.~\cite{CavFucSch02a}) 
should neglect the epistemic component to a
state like that in  Eq.~(1). 

In fairness to Fuchs' position, I should
reproduce a further argument 
he has communicated to me \cite{Fuchs:pc}{}.
It is based on a theorem 
discovered by \sch \cite{Sch36} and called by him
``distant 
steering" (see also Ref.~\cite{VujHer88}). This theorem states
that if a system is in a mixed state by virtue of being entangled (that
is, there is a pure state in the larger Hilbert space of the system 
and its environment) then there is a way to measure the environment 
such as to
``steer" the system into any one of the ensembles of pure 
states that
represents that mixed state. Since the average state of 
the system is
unchanged regardless of which ensemble is used to 
represent it, it is
arguable as to whether steering is an appropriate 
term. Nevertheless its
application in this case is clear. If the 
laser used by Alice and Bob for
CVQT is in a mixed state purely by virtue of being entangled with its
environment, then there would be a way to for a third independent
observer (call him Chris) to measure that environment so as to 
collapse
the laser into a number state $\ket{n}$, with $n$ randomly 
chosen from the
probability distribution in Eq.~(2). If Chris knows 
the laser to be in
state $\ket{n}$, then how can Alice and Bob claim 
that it, the clock, is
in a coherent state $\ket{\sqrt{\mu}}$?

One may well imagine the response
of Alice and Bob to Chris' attempt 
to carry out this measurement: ``get
your filthy hands off our 
clock!" By measuring the environment so as to
collapse the laser into 
a number state, Chris would prevent it from being
used for its 
desired purpose, namely to be a clock. Chris would have to
detect all 
of the light emitted by the laser, so there is no way Alice and
Bob could use it to establish a phase reference. Even if Chris did a
quantum-non-demolition measurement of the photon number of the light,
this would destroy its phase and so still prevent it from being used 
to establish a definite phase between Alice and Bob. It is not 
surprising that, when it cannot be used as a clock, it is not valid 
to claim that a
laser has a well-defined phase. But when a laser is 
being used as the
clock for the experiment there is no way Chris can 
know which number state
it is in, and no reason for Alice and Bob not 
to describe it by a coherent
state $\ket{\sqrt{\mu}}$. 

To sum up my position: a quantum state is not
only a state of knowledge, it is a state of purpose. That is, it is not
only what I know about a system, but also how I intend to use it, that
determines how it is described  mathematically. To ignore this fact is to
take the mathematical formalism of quantum mechanics too seriously.
Operationally, sharing a laser is simply clock synchronization. 
There is no puzzle here, and our 
formalism should reflect that. To argue otherwise
may be interesting 
philosophy, but it is not physics (at least as defined
by the {\em Physical Review}, despite the evidence
\cite{RudSan01,EnkFuc02a} to the contrary).

\section{Objection: Conservation of energy implies a superselection rule for photon number}
 
A super-selection rule (SSR)
for a particular observable $\hat{O}$ 
is a restriction on the possible
operations that can be implemented 
physically. Roughly, it says that it is
impossible to produce states 
that are superpositions of eigenstates of
$\hat{O}$ with different 
eigenvalues. For a more precise exposition see
Ref.~\cite{BarWis03}{}.

Neither RS, nor EF, nor I, claim that there is a
SSR for photon number. For RS, and EF, this is because they believe that
there is a way, with a classical field (RS), or a microwave field (EF), to
produce such optical coherences. For my part, I claim that whether or 
not a phase is a superposition or a mixture depends upon how it is 
used. If it
is used as a clock, then it is coherent by definition. 
However, the
notion that conservation of energy implies a SSR for 
photon number does
exist in the community. If it were true this would 
present a problem for
my argument, and in fact would be a physical 
basis for preferring the
ensemble in Eq.~(1) to that in Eq.~(2). 
Therefore in this section I will
rebut that notion. 

To begin it is necessary to clear up some
misconceptions. First clarification: {\em SSRs should be applied only to
non-conserved quantities}. That is not to say that SSRs are unrelated to
conserved quantities, but the relation is more subtle, and I have not seen
a decent exposition elsewhere. Here is mine. It is obvious that one
cannot create superpositions of different values of conserved 
quantities. Conserved quantities do not change, by definition. If 
initially the system
were in an eigenstate of a quantity  then it 
would of course be impossible
to create a superposition, because this 
would be equivalent to changing
$\hat{O}$. The same is true for 
initial mixtures. Thus there is no need to
invoke an SSR to say that 
creating superpositions is impossible. The
conservation law already 
tells one this. The only useful SSR is one that
applies to 
non-conserved quantities.

The relation between SSRs and
conservation laws is as follows. 
If  and only if a conserved quantity
$\hat C$ can be written as
\beq
\hat C = \hat c_1\otimes \hat 1_2
\otimes\hat 1_3 \cdots 
\otimes \hat 1_N \,+\,
\hat 1_1 \otimes \hat c_2
\otimes \hat 1_3 
\cdots \otimes \hat 1_N \,+\,\, \cdots \,\,+\,
\hat 1_1
\otimes \hat 1_2 
\otimes  \cdots  \hat 1_{N-1} \otimes \hat
c_N,\label{sum}
\eeq
then it is possible to derive a SSR (in fact, $N$ of
them). Namely, 
these are that it is 
impossible to produce superpositions
of states having different 
eigenvalues of $\hat{c}_k$, for $k = 1\cdots
N$. This follows from 
the presumption that the initial state (of the
universe, or of the 
laboratory if one wishes to be more modest) is an
eigenstate, 
 of $\hat{C}$. Note that $\hat{c}_k$ is {\em not} in general
conserved. Nevertheless it obeys a SSR, because if a state were a
superposition 
of different eigenstates of $\hat{c}_1$, for example, then
by the structure of \erf{sum} it would be a superposition of different
eigenstates of $\hat{C}$, and that is forbidden by the initial 
conditions and the conservation of $\hat{C}$. 
 
An example may help. {\em Total
momentum} $\hat{P}$ is a conserved 
quantity. Furthermore, total momentum
is composed of a sum of momenta 
of individual particles and field modes,
as in \erf{sum}. These 
individual momenta $\hat p_k$ are not conserved, as
the particles and 
fields interact. But, if the universe were to have begun
in a momentum eigenstate, then there would be a SSR for every individual
momentum. That is, no particle could be localized, because it would
always be a mixture of momentum eigenstates, never a superposition. 
This is simply because the initial state of the universe would be 
completely
delocalized also.

Perhaps this is a poor example, because saying that the
universe began 
in an eigenstate of momentum  is arguably a meaningless
statement. This is because it is meaningless to discuss a spatial
reference frame in the absence of the matter of the universe, just as 
it is meaningless to discuss the phase of an oscillator in the 
absence of a
clock. Thus, 
one could argue that relative to itself the universe is not
in a momentum eigenstate, 
since it is clearly somewhere definite, namely
exactly on top of 
itself. The same argument would hold for 
spatial orientation, and the corresponding SSR for angular momentum. 
By contrast,
claiming that the universe is in an eigenstate of total 
charge is perhaps
meaningful, at least in some cosmologies. Since 
total charge is also the
sum of charges of subsystems, a SSR for the 
charge of such systems (e.g. a
certain volume of space) is arguably 
better justified.

Regardless of
whether this derivation of SSRs from conservation laws 
and initial
conditions is ever justified, it is certainly {\em not} 
justified for
energy. The total energy $\hat{H}$ of the universe may 
be conserved, but
it is not of the form of \erf{sum}. 
 If it were of this form then that
would mean that the universe would consist of some disjoint parts which
have no interaction whatsoever with one another. So really, the
Hamiltonian 
$\hat{H}$ would be describing a number of universes, not one
universe. Thus, 
there is no energy SSR, so a photon number SSR cannot be
derived from it.

In the quantum optical case, it is the interaction
Hamiltonian between the electromagnetic field and charged particles that
prevent the Hamiltonian being of the form of \erf{sum}. This is not an
inconvenience that might be ignored. Rather, is crucial to the fact 
that an electromagnetic field can be a clock. As noted in Sec.~II, it 
is the ${\bf p}\cdot{\bf A}({\bf x})$ coupling that makes the 
electromagnetic
phase reference fungible with all other clocks. 
 
\section{Objection: A
laser beam needs a quantum channel}
 
Even if one accepts all of my above
arguments, one may still wonder 
about the initial issue which at least
partly motivated RS, namely that the 
laser beam is acting as a quantum
channel. Operationally this problem 
could be easily removed by defining
quantum teleportation to require 
that the phase reference channel, be it
quantum or channel, must not 
be open during the teleportation itself. That
is, setting up the time 
reference would, like the sharing of entanglement,
be part of the 
setup prior to teleportation. This would certainly be
possible in 
principle although in practice it would be inconvenient as it
would 
require maintaining the optical phase of a light field (such as that
in a high-finesse cavity) for a relatively long time (perhaps many
$\mu$s) on each side. 

However, it is interesting nevertheless to ask the
question, is a 
quantum channel necessary to share a phase reference? To
answer this 
one needs a way to distinguish a  a classical and a quantum
channel. 
The obvious definition is that used, perhaps for the first time,
in Ref.~\cite{BarNieSch98} : a classical channel is a channel that does
not permit the sending of quantum information. Specifically, a 
classical
channel completely decoheres the system in some basis. That 
is, it
diagonalizes the state matrix in that basis, suppressing the 
coherences.

Since the coherent states are non-orthogonal and overcomplete:
\beq
\int
d^2\alpha \ket{\alpha}\bra{\alpha} = \pi \hat{1}
\eeq
it might be thought
that it would be impossible to send coherence 
down a quantum channel. It
is true that it is impossible to send a 
coherent state down a classical
channel and have it emerge unscathed. 
However, to send coherence (i.e. a
phase reference) this is not 
necessary. All that is necessary is for the
emerging state to have a 
large coherent amplitude which is approximately
equal to that of the 
coherent state that entered. This is quite possible.

Consider as a crude example the states $\cu{\ket{q_n,p_m}:n,m 
\textrm{
integers}}$ defined in the $\hat{q}$-representation as
\beq
\ip{q}{q_n,p_n} =
\Delta^{-1/2}\chi_{[q_n-\Delta/2,q_n+\Delta/2]}(q)\exp(-iqp_m).
\eeq
Here
$q_n = \Delta n$ and $p_m = 2\pi m/\Delta$, $\chi_S(x)$ is the 
characteristic function, equal to $1$ when $x\in S$ and $0$ 
otherwise, 
and  $\hat{q} = (
a+a\dg)/\sqrt{2}$. 
These states have a mean coherent
amplitude:
\beq
\bra{q_n,p_m}{a}\ket{q_n,p_n} =
(q_n+ip_m)/\sqrt{2}.
\eeq
The variance (and indeed all moments) are finite
for $\hat{q}$. For 
$\hat{p} = -i(a-a\dg)/\sqrt{2}$ the variance  (and
indeed all 
moments) are infinite. Nevertheless the probability
distribution 
$P(p)$ is peaked (with a width of order $\Delta^{-1}$) about
$p=p_m$. 
Thus with a choice of $\Delta \sim 1$ this is a state with
well-defined coherent amplitude, and a fair overlap with a coherent 
state
$\ket{\alpha}$ with $\alpha$ close to $(q_n+ip_m)/\sqrt{2}$. 

The states
$\cu{\ket{q_n,p_m}:n,m}$ form an orthonormal set with the 
completeness
relation 
\beq
\sum_{n,m} \ket{q_n,p_m}\bra{q_n,p_m} = \hat{1}.
\eeq
Thus
they can be used to define a classical channel, by specifying 
the
trace-preserving operation on the channel
\beq \label{decop}
\rho \to {\cal
O} \rho = \sum_{n,m = -\infty}^\infty  
\ket{q_n,p_m}\bra{q_n,p_m} \rho
\ket{q_n,p_m}\bra{q_n,p_m} 
\eeq
Clearly a pure coherent state
$\ket{\alpha}$ will become mixed by 
this channel. However,   as long as it
is a good phase reference, so 
that $|\alpha| \gg 1$, the state that
emerges 
\beq
\sum_{n,m}   |\ip{q_n,p_m}{\alpha}|^2
\ket{q_n,p_m}\bra{q_n,p_m}  
\eeq
will be dominated by states with
$(q_n+ip_m)/\sqrt{2} \simeq 
\alpha$.  Thus the state still carries the
phase information 
necessary to establish a phase reference. 

It might be
objected that in the experiment there is no channel which 
is described by
the operation (\ref{decop}), and thus that in the 
experiment there is a
quantum channel, not a classical channel. But 
this seems unfair. Say an
experimenter E claims to be using a channel 
as a classical, not a quantum,
channel. Say a skeptic S challenges 
the validity of an experiment by
claiming that actually it is being 
used as a clandestine quantum channel.
E comes back and says, ``Okay, 
to prove it is being used as a classical
channel, I will repeat the 
experiment while allowing you (S)  to decohere
the channel in this 
basis which I specify." S says ``But I don't know how
to make that 
operation." Says E, ``Tough -- in a duel the one who is
challenged 
gets to choose the weapons. Here the onus is on you if you want
to invalidate my experiment."

Following on from this argument, the
stymied challenger will (as I 
well know) continue to raise the objection
that without the decohering 
mechanism, an optical channel (e.g. free space
or a fiber) {\em can} 
be used to send quantum information. In a genuinely
classical way to 
synchronize clocks, this would not be the case. For
example, one 
could send marbles down a pipe so that they arrive at the far end
at regular times, separated by $10^{-15}$s. 

The error with this argument
is that it assumes that, just because we 
ordinarily describe marbles as
classical objects, that they really 
are classical objects. If one allows
for the existence of classical 
objects by fiat, then by fiat I would claim
that a laser beam is a 
classical object. Certainly to an ordinary laser
physicist, as 
opposed to a quantum optician, laser beams {\em are}
classical objects, and the former would laugh at the idea that
synchronizing 
clocks using laser beams was sending quantum information.

Allowing instead that marbles are quantum objects, how is a free 
space channel used for clock-synchronization by marble arrival-time a 
quantum
channel? The answer is in the longitudinal momentum modes of 
the marbles.
If the marbles arrive at definite times (as was the 
original assumption)
then  they must have well-defined wave-packets 
in position space. That is,
they must be superpositions of different 
momenta. In principle, this could
be used to encode quantum 
information, as the superposition of momentum
eigenstates is 
preserved by free evolution. 

Just as in the laser case,
it would be possible to defend the marble 
channel as a classical channel,
by  choosing some basis set of 
orthogonal states along which to decohere
the channel. These could 
not be momentum eigenstates, as this would make
the position of the 
marbles completely delocalized, so all timing
information would be 
lost. Nor could it be position eigenstates, as these
are states of 
infinite energy which would immediately become delocalized
also. Rather, just as in the laser case, what would be required is states
localized in phase space with some suitably chosen position width
$\Delta$. 

No one would take you seriously if you insisted that an
experiment 
using marble-timing to synchronize clocks is really using
quantum 
information, unless the experimenters make sure they decohere the
longitudinal quantum state of their marbles completely along some 
basis. The point is that there is absolutely no difference using a 
laser. Exactly
the same issues will always arise, if one thinks 
carefully, regarding the
preparation and dissemination of states 
which can be used to synchronize
clocks,  at least for any frequency 
higher than that which can be
consciously controlled or perceived.
 
 \section{Objection: A laser cannot
synchronize arbitrarily many 
parties}

The final objection I will consider
is that  a laser beam is not a 
clock because it cannot 
establish a time
standard between arbitrarily many parties. 
Eventually 
it will run out,
or, if it is a continuous beam, its finite linewidth 
will mean that the
later part of the beam has a random phase relative 
to the former part.
This is of course true, and the fundamental 
limits are set by the
finiteness of the excitation of the laser mode and laser gain medium
\cite{Wis99b}{}. This excitation may be very 
large (measured in units of
$\hbar \omega$), but is not infinite. 

Once again, the mistake in this
reasoning is to ignore the quantum 
nature of 
other clocks, simply because
we are used to describing them 
classically. All physical clocks have
finite excitation, and cannot 
be used to synchronize arbitrarily many
parties.
Low frequency material clocks typically have huge excitations (in
units of $\hbar \omega$), so we are not used to worrying about their
finiteness. A similar point has been made in
Ref.~\cite{EguGarRay99}{}.

At optical frequencies, a laser is actually
the best clock we have. 
In fact, at
the National Institute of Standards in
Boulder, Colorado,
the next generation of an "atomic clock" in
development
is actually an optical clock using a laser \cite{Did01}{}. As
pointed 
out in Ref.~\cite{HeyBjo03}{}, this is a powerful argument in
support 
of my view that 
a laser already produces a coherent state. Once a
laser becomes the 
basis for the world's time standard, then establishing
the phase of a 
laser relative to a mechanical oscillator will fix the
phase of the 
{\em latter}, rather than the {\em former}. 

Nevertheless,
it is an interesting question to ask precisely how good 
could a laser be
as a time standard? That is, what are the quantum 
limits to how many
clocks could be simultaneously synchronized with 
the standard laser.
Clearly for an infinitely large laser the answer 
is that there is no
limit. But for a laser with a finite excitation, 
there must be a definite
answer, and to my knowledge, it has not been 
given before. The answer is
that for a laser with a mean excitation 
number $\mu$, the number $M$ of
clocks that can be synchronized with 
a mean-square error $(\delta\phi)^2$
(in radian$^2$) satisfies
\beq \label{Heislim}
(\delta\phi)^2 \geq
\sqrt{M}/4\mu
\eeq
This is, to me at least, not an obvious expression, so
readers may 
wish to see if they can derive it (ignoring factors of order
unity)  
from simple arguments, and to inform me if they can.

The first
point to note is that this expression cannot be derived 
simply by
considering the dividing of a coherent state of amplitude 
$\sqrt{\mu}$. In
the large-amplitude limit, the phase variance of a 
coherent state
$\ket{\alpha}$ is equal to $|\alpha|^{-2}/4$ (see e.g.
Ref.~\cite{WalMil94}). Thus if the coherent state $\ket{\sqrt{\mu}}$ 
were
split coherently into $M$ coherent states of amplitude 
$\sqrt{\mu/M}$, the
phase variance of each one individually would be
\beq
(\delta\phi)^2 \sim M/4 \mu
\eeq
Note the difference from the above by a factor of $\sqrt{M}$.
For $M$ 
large, which is the case we are interested in, this phase variance
may be thought of as the mean square phase error relative to the 
original
coherent state $\ket{\sqrt{\mu}}$ which has negligible phase 
uncertainty
itself. However these small coherent states may be 
amplified, the phase
error relative to the original cannot be reduced.

The difference between a
laser with mean photon number $\mu$ and a 
coherent state with mean photon
number $\mu$ is that the laser 
maintains its number even as it generates
an output that can be 
shared amongst many parties. The price paid by this
is that the phase 
of the laser does not remain constant over time. Rather,
it diffuses. 
To derive \erf{Heislim} it is necessary to answer two
questions. 
First, what is the quantum limit to the rate of laser phase
diffusion? Second, what is the quantum limit to how well a second 
clock
can be synchronized to a continuous signal with a diffusing 
phase? That is
the content of the next two subsections.

\subsection{Quantum limits to
laser phase diffusion.}

This topic will in most readers' minds conjure up
the names of 
Schawlow and Townes who introduced the 
idea of an ``optical
maser'' some 45 years ago \cite{SchTow58}{}. 
Actually, the expression in
their paper for the quantum-limited laser 
linewidth is irrelevant for most
modern lasers. That is because it 
was derived in the days 
before good
optical cavities, 
and hence implicitly assumes that the atomic linewidth
$\gamma$ is 
much smaller than the (FWHM) cavity linewidth $\kappa$. 
This is discussed in detail in Ref.~\cite{Wis99b}{}. 

In more recent times, the
term ``Schawlow-Townes limit'' has commonly 
(including by myself
\cite{Wis97}), although arguably erroneously, 
been used to refer to the
quantum limit for the linewidth of a laser 
in the limit that the atomic
medium can be adiabatically eliminated 
(which requires $\kappa \ll
\gamma$). Far above threshold, and 
ignoring effects such as finite
temperature and reabsorption, this 
linewidth $\ell$ is
\beq
\label{ell0}
\ell_{\rm SQL} = \frac{\kappa}{2{\mu}}.
\eeq
 Here $\mu$ is,
as above, the mean photon number of the laser.
 I will call this the
standard quantum limit (SQL) linewidth. As I 
will discuss below, it is
{\em not} the ultimate limit to the 
linewidth set by quantum mechanics,
which I will call, following 
standard practice in other areas, the
Heisenberg limit (HL) linewidth.

Most older textbooks
\cite{Lou73,SarScuLam74,Lou83} quote the result 
in \erf{ell0}, 
or one
which reduces to it in the appropriate limit of neither 
reabsorption 
nor
thermal photons. All three of these books attempt to explain this
linewidth in terms of 
the noise added by the spontaneous contribution to
the (mostly 
stimulated) gain of photons from the atomic medium. Far above
threshold the amplitude of the laser is essentially fixed. Thus the
effect of spontaneous emission noise is (supposedly)  to randomly 
perturb
the laser phase (relative to some hypothetical time 
standard). This can be
modeled by the equation
 \beq \label{phasediff}
 \dot{\phi} =
\sqrt{\ell}\xi(t),
 \eeq
 where $\xi(t)$ is Gaussian white noise satisfying
$\an{\xi(t)\xi(t')} = \delta(t-t')$. This gives rise to a Lorentzian
lineshape of linewidth $\ell$. 
 
 It turns out \cite{Wis99b} that these
explanations are strictly 
incorrect. Attributing the phase diffusion
solely to the gain 
mechanism is an artifact 
of thinking in terms of
normally-ordered operator products. 
That is, it results from using
(implicitly in most cases) the 
Glauber-Sudarshan $P$ function
\cite{WalMil94} 
as a true representation of the 
the fluctuations in the
laser mode field. The $P$ function is of 
course no more fundamental than
the $Q$ function \cite{WalMil94}{}, 
which is a 
representation based on
anti-normally 
ordered statistics. If one were to use  
the $Q$ function as
an aid to intuition, one would find that it is 
the loss process that is
wholly responsible for the phase noise. Of 
course the rate of phase
diffusion would agree with that from the $P$ 
function, at least in
steady-state where loss and gain balance. 

If one asks a question about
phase diffusion, the only objective 
answer will come from using the phase
basis itself. This is far more 
difficult than using the more familiar
phase-space representations, 
but some approximate results have been
obtained \cite{BarStePeg89}{}. 
These show 
that, at steady state, the
phase diffusion has equal contributions 
from the loss and gain process.
The same result occurs from a Wigner 
function calculation
\cite{BarStePeg89}{}. This is not surprising 
since symmetrically 
ordered
moments are known to closely approximate the true moments 
for the phase
operator for states with well-defined amplitude 
\cite{HilFreSch95}{}.

The fact that standard phase diffusion comes equally from the loss 
and gain 
processes suggests that the SQL to the laser 
linewidth, $\ell_{\rm
SQL}$ of \erf{ell0}, may not be the ultimate 
quantum 
limit. The
contribution from the loss mechanism is unavoidable. A 
laser, at least in
useful definitions~\cite{Wis97}{}, requires  
linear damping of the laser
mode (here with decay rate $\kappa$) in 
order to form an output beam.
However, it may be that the standard 
gain mechanism could be replaced 
by some other gain mechanism that still yields the same steady state 
(1), but
which causes less phase diffusion. That is, 
we assume that the state
matrix $\rho$ for the laser mode obeys the 
master equation
\beq
\dot{\rho}
= {\cal L}_{\rm gain}[a,a\dg] \rho + \kappa (a\rho a\dg - 
\half \{a\dg
a,\rho\}) -i\omega[a\dg a,\rho]
\eeq
where the gain term is
time-independent (that is, involves no 
external driving), but may be
different from the standard laser gain 
term. The HL to the laser linewidth
from such an equation could be as 
small as one half of the standard
limit.
 
 It turns out that it is indeed possible to find a gain
Liouvillian 
${\cal L}_{\rm gain}[a,a\dg] $ which, contrary to the standard
term, 
adds no phase noise  \cite{Wis99b}{}. For interest, it is given
explicitly by 
  \beq \label{nsgain}
{\cal L}_{\rm gain}[a,a\dg]\rho =
\kappa\mu[
a\dg (a a^{\dagger})^{-1/2} \rho (a a^{\dagger})^{-1/2} a
-\rho].
\eeq
Moreover, there is physical mechanism by which this gain could
be realized, using adiabatic transitions in cavity QED \cite{Wis99b}{}.
The HL laser linewidth is thus half the SQL:
\beq \label{ellHL}
\ell_{\rm
HL} = \frac{\kappa}{4{\mu}}.
\eeq

\subsection{Quantum limits to phase locking}

From the above, we know that we can model the output of a laser
as a 
coherent state with a phase stochastically evolving according to
\erf{phasediff}. The power of the laser output beam is $P = 
\hbar\omega
\kappa \mu$. It is more convenient to measure power in 
photons per second,
for which I will use the symbol $p$. If the laser 
output is split equally
between $M$ parties, for the purposes of 
clock synchronization, then the
power in each beam is 
$p=\kappa\mu/M$. 

The question is then, what is the
best way to synchronize one's own 
clock (e.g. a laser) to a continuous
beam of power $p$ with a phase 
diffusing according to \erf{phasediff}. I
will assume that each party 
has their own laser, with an arbitrarily large
photon number, so that 
the rate of phase diffusion of these lasers is
negligible compared to 
the rate of phase diffusion $\ell$ of the source
laser (to which all 
others are synchronized). In practice, one would
expect the reverse 
situation; that is, one would want the standard laser
to have the 
lowest noise. However for the purposes of determining the
limit to 
how good a given, finite, laser can be as a phase reference, it
is necessary to consider the above situation.

One possibility for clock
synchronization, considered in 
Ref.~\cite{HeyBjo03}{}, is to use the beam
as a seed to one's own 
laser. As Heydar and Bj\"ork show, this is
certainly adequate, but 
they do not consider its quantum limits. Since the
gain term in a 
standard laser does introduce unnecessary phase noise (as
discussed 
above) it seems unlikely that this process would be the optimal
one. It is possible that a laser with a non-standard gain term, as in
\erf{nsgain}, would be optimal, but the analysis of that question is
beyond the scope of this paper.

Rather than using the source beam to seed
the local laser, the phase 
of the latter can be locked to the former by
measuring the phase 
difference between them, and adjusting the phase of
the latter so as 
to eliminate as far as possible this phase difference.
This is 
precisely the problem considered in Ref.~\cite{BerWis02}{}: to
estimate continuously the phase of a CW beam with a finite power and
non-zero linewidth (i.e. finite phase diffusion rate). An elementary
discussion of this problem is given in Ref.~\cite{Wis02c}{}.

It turns
out, unsurprisingly \cite{Wis02c}{}, that the optimal 
technique for
estimating the phase using conventional experimental 
techniques (linear
optics and photodetection) is an adaptive one. 
That is, it is necessary to
use past measurement results 
(photocurrents) to control the present
conditions of the measurement 
in order to obtain the maximum information
about the phase. Attaining 
the \hei\ limit of phase locking requires quite
precise control of the 
feedback gain the in the locking loop. 

For coherent light (i.e. non-squeezed light) as we are considering 
here, using
a {\em non-adaptive} technique, the mean square error in 
the best estimate
for the phase $\phi$ is, in steady state, 
\cite{BerWis02}
\beq
(\delta\phi)^2_{\rm SQL} = \frac{1}{\sqrt{2N}}
\eeq
Here $N = p/\ell$ is a
dimensionless parameter which (up to factors 
of order unity) is equal to
the number of photons in the beam per 
coherence time, or, equivalently,
the peak photon current per unit 
bandwidth. We have assumed $N\gg 1$
because this is necessary to 
obtain good clock synchronization. Note that
the phase variance 
scales as $N^{-1/2}$, not $N^{-1}$ as might be naively
expected.

 An example of a non-adaptive measurement technique is
ambi-quadrature homodyne detection, where the beam is split in two, 
with
one half being used to perform a measurement of one quadrature 
and the
other half used to perform a measurement of the orthogonal 
quadrature.
This non-adaptive technique is wasteful because it 
obtains information
about the amplitude equally as well as about the 
phase, of the beam.
Because these are complementary quantities, 
obtaining the amplitude
information halves the amount of phase 
information that can be obtained. 

 An adaptive technique ensures that ones obtains, at least 
approximately,
only phase information. As in the discussion in the 
preceding subsection
on laser gain mechanisms, the boon granted by 
eliminating unnecessary
phase uncertainty is modest. Specifically, 
the adaptive technique yields
\cite{BerWis02,Wis02c}
\beq \label{adapt}
 (\delta\phi)^2_{\rm adapt.} =
\frac{1}{\sqrt{2N}}.
\eeq
Note that I have not labeled this the \hei\
limit. That is because a 
phase-squeezed source can actually do better,
with a scaling of 
$N^{-2/3}$, as derived in Ref.~\cite{BerWis02}{}. There
is no point 
considering that here, because we are interested in the case
where 
the beam to which we are locking arises from a laser. A
phase-squeezed beam could only be produced by starting with a much 
more
powerful source laser, and using a non-linear crystal for 
example. Such a
process could be useful if we were interested in 
limiting the power in
each individual beam, $p$. But here we are 
interested only in the ultimate
limit set by the photon number in the 
source laser, $\mu$.

\subsection{Quantum limits to the laser as a clock}

We are now in a
position to combine the results of the preceding two 
subsections. Using
the \hei\ limited laser linewidth (\ref{ellHL}), 
and the optimal (adaptive)
result for phase locking  (\ref{adapt}) 
with $N = \kappa \mu / M\ell$, we
obtain the HL
\beq \label{HL}
 (\delta\phi)^2_{\rm HL} =
{\sqrt{M}}/{4\mu},
\eeq
as quoted in \erf{Heislim}. Using a standard
(ideal) laser, and a 
non-adaptive technique, one would obtain the
SQL
\beq
 (\delta\phi)^2_{\rm SQL} = {\sqrt{M}}/{2\mu},
\eeq
only a factor
of two greater. 

In practice, the laser linewidth $\ell$ is usually not
limited by the 
uncertainty principle, and is far greater than
$\kappa/2\mu$. The 
general expression, which may actually be of use to
experimentalists, 
is
\beq
(\delta \phi)^2 \sim \sqrt\frac{\hbar \omega M
\ell}{P}.
\eeq
For example, a one mW laser with a linewidth of 1 MHz in the
visible 
can synchronize the clocks of $M$ parties with a mean square error
of
\beq
(\delta \phi)^2 \sim 10^{-5} \sqrt{M}
\eeq
Thus, every person on
the planet could use this laser independently 
to synchronize their clocks,
with an error of order $1$ radian (that 
is, of order $10^{-15}$s). 

Of course, this would not be the most efficient way to establish an
optical-scale time standard among the world's population. Rather, if 
the
above, very ordinary, laser had to be used as the time standard, 
then it
would be best to lock it to another laser, with far better 
properties
(higher power, lower linewidth) and then to use that laser 
as the new
standard. 
Providing the second laser was good enough, this would result in
no further increase in mean square phase error, relative to the original
source laser,  above the  $10^{-5}$ radians$^2$ between the source 
and second laser. This reinforces the idea that the phase reference 
supplied
by the source laser is not a quantum channel. If used 
wisely, the
information can be shared amongst many parties without 
further loss. Thus
it makes sense to think of it as classical 
information, not a quantum
resource.

\section{Conclusion}

I have argued that, even though it
is correct to represent the 
quantum state of a laser by a mixture of
coherent states (1) or 
number states (2), the output of a laser is truly
coherent in any 
meaningful sense of the word. In particular, it is also
quite correct 
to represent the state as a single coherent state
$\ket{\sqrt{\mu}}$. 
That is because the state of any oscillator, except
for an energy 
eigenstate, can only be defined {\em relative} to some time
reference. If a laser itself is used as the time reference, then by
definition it is coherent relative to itself, with phase zero. Thus, 
the
idea of Rudolph and Sanders, accepted also by van Enk and Fuchs, 
that
there is some `true coherence' which a laser lacks (as yet) is 
illusory.
Thus there is no sound basis for the former's criticism of 
the continuous
variable quantum teleportation experiment 
\cite{Fur98}{}.

In defending
this view I have made a number of important points. 
First, absolute
phase, at least on any time scale beyond that  of direct human 
experience
($\sim 10^{-1}$s), is meaningless. There is only relative 
phase, and a
laser is as good a phase reference as any others. 
Indeed, a laser deserves
to be called a clock because its time 
standard is {\em fungible} with any
other clock. The same is not true 
of a Bose condensate, for example.

Second, even though Eq.~(2) implies that a sufficiently clever 
observer
could find out which number state a laser is in, this does 
not contradict
my statement that a laser used as a clock is in a 
coherent state. That is
because finding out which number state a 
laser occupies would destroy the
phase information necessary for it 
to be used as a clock. A quantum state
is a state of purpose as well 
as a state of knowledge.

Third, the
existence of optical phase cannot be denied on the basis 
of a
super-selection rule for energy, because no such super-selection 
rule
exists. I have shown this by carefully explaining that a 
super-selection
rule relates to a conserved quantity if and 
only if it is of the form of a
direct sums of other quantities, and 
that the rule applies to these other,
{\em non-conserved} quantities. 
Energy is never of this form.

Fourth, it
is not necessary to have a channel which can transmit 
quantum information
in order to establish a time reference, using a 
laser or any other means.
Clock synchronization in the teleportation 
experiment can thus be regarded
as quite separate from the quantum 
teleportation.

Fifth, although a given
laser can only synchronize several parties to 
a finite degree of accuracy,
this is not an argument against it being 
a clock. That argument would be
true of any  clock with a finite 
excitation. In this paper I have
determined, to my knowledge for the 
first time, the quantum limits to
clock synchronization using a 
laser. The result, which at least to me was
not obvious, is that the 
accuracy with which the phase of $M$ parties can
be independently 
locked to that of a continuously running laser containing
on average 
$\mu$ quanta of excitation is limited by 
\beq
(\delta \phi)^2
\geq {\sqrt{M}}/{4\mu}.
\eeq

None of my arguments strictly disprove the
existence the existence of 
absolute time, even if it is unobservable. So,
if one wishes, one 
could insist that the laser should always be described
by Eqs. (1) or 
(2) rather than a single coherent state. One could follow
Rudolph and 
Sanders and insist therefore that CVQT has not been
demonstrated. But 
one should be aware that by doing this one must also
insist that 
there would never be any states with well-defined phase to be
teleported either. And that no experiment has ever demonstrated 
optical
squeezing, let alone optical entanglement. In fact, their 
arguments would,
carried to their 
logical conclusion, banish 
from our theories any time
$t$ or phase $\phi$ if its 
implied resolution were beyond that of direct
human experience. 
To scientists and engineers, this would be unacceptable
pedantry.


\section*{ACKNOWLEDGMENTS}       
 
I gratefully acknowledge friendly
discussions with S. Bartlett, S. 
Braunstein, C. Fuchs, T. Rudolph, B. 
Sanders, and R. Spekkens.


\end{document}